\documentclass[reprint,aps,pra,prb]{revtex4-1}
\usepackage{graphicx}
\usepackage{dcolumn}
\usepackage{bm}
\usepackage{amsmath}
\usepackage{amsthm}
\usepackage{amssymb}
\usepackage{float}
\usepackage{hyperref}
\usepackage{color}
\usepackage{epstopdf}
\usepackage{cleveref}
\usepackage[svgnames]{xcolor}
\hypersetup{hidelinks,colorlinks=true,allcolors=DarkBlue}

\begin{document}

\preprint{APS/123-QED}

\title{Roton-maxon spectrum and instability for \\ weakly interacting dipolar excitons in a semiconductor layer}
\author{A.\,K.\,Fedorov$^{1}$}
\author{I.\,L.\,Kurbakov$^{2}$}
\author{Yu.\,E.\,Lozovik$^{2,3,4}$}
\affiliation
{
\mbox{$^{1}$Russian Quantum Center, Skolkovo, Moscow 143025, Russia}
\mbox{$^{2}$Institute of Spectroscopy, Russian Academy of Sciences, Troitsk, Moscow Region 142190, Russia}
\mbox{$^{3}$Center for Basic Research, All-Russia Research Institute of Automatics, Moscow 127055, Russia}
\mbox{$^{4}$MIEM at National Research University HSE, Moscow 109028, Russia}
}
\date{\today}

\begin{abstract}
The formation of the roton-maxon excitation spectrum and the roton instability effect for a {\it weakly} correlated Bose gas of dipolar excitons in a semiconductor layer are predicted.
The stability diagram is calculated. 
According to our numerical estimations, the threshold of the roton instability for Bose-Einstein condensed exciton gas with roton-maxon spectrum is achievable experimentally, {\it e.g.}, in GaAs semiconductor layers.

\begin{description}
\item[PACS numbers]
71.35.Lk, 03.75.Hh, 02.70.Ss, 73.21.Fg
\end{description}
\end{abstract}

                              
\maketitle

\section{Introduction}
Experimental observation of Bose-Einstein condensation (BEC) of ultracold diluted atomic gases has opened new frontiers in physics \cite{BECat}. 
Interesting phenomena have been predicted and observed in atomic condensates \cite{atomrev}. 
The BEC critical temperature is inversely proportional to the effective particle mass.
Therefore, investigations of the collective properties and BEC of excitons, whose effective mass is much smaller not only atomic mass but even electron mass, are highly promising 
(see \cite{KK, Moskalenko, B} and references therein). 

In general, the lifetime of excitons is not long enough to achieve the thermodynamical equilibrium. 
To overcome this difficulty, excitons with spatially separated electrons ($e$) and holes ($h$) can be used \cite{LN,LY,Neilson,CD}. 
This separation suppresses the recombination process, {\it i.e.}, the lifetime of excitons significantly increases.
Several realizations with increased lifetime have been proposed such as 
(i) two-dimensional (2D) semiconductor heterostructures with two coupled quantum wells (QWs) separated by a barrier \cite{LY}; 
(ii) a single QW in normal to its plane electric field \cite{K,F};
(iii) independently gated two graphene monolayers separated by a hBN insulating barrier \cite{graphene,graphene2}; 
(iv) independently gated opposite surfaces with 2D Dirac systems on thin films of topological insulators \cite{TI,TI2};
(v) MoS$_2$ layers separated by an hBN insulating barrier and surrounded by hBN cladding layers \cite{FBN}.
For systems with spatially separated $e$ and $h$, the BCS-type regime \cite{LY,PNS}, BEC \cite{LY2} as well as the BCS-BEC crossover have been considered \cite{LB}.

Furthermore, the $e$--$h$ separation results in the appearance of the exciton dipole moment. 
Dipolar excitons for coupled QWs \cite{LY} and for single QW in electric field \cite{F} have been studied. 
In sufficiently thin QWs, systems of dipolar excitons have 2D behavior, because dynamics of dipolar exciton systems is quantized in the tight direction, and dipole moments of all excitons are equal and normal to the QW plane. 
Persistent currents \cite{LY}, superfluidity \cite{LB2,SF}, Josephson-like effect \cite{JLE}, anomalous optical properties \cite{opt}, 
and spin effects \cite{prb88195309, B2, prl109026401,prb2014,jetp11900115} of 2D dipolar excitons have been considered. 
Outstanding findings \cite{EE,LO} have been reported on exciton BEC in $e$-$e$ bilayer in strong magnetic fields. 
Interesting effects in the collective state of 2D dipolar excitons in QWs have been observed experimentally \cite{B,B2,jetp11900115,M,T,Snoke,nc0004002335,Stern}  (see also related works \cite{Dubin}). 

\begin{figure}
\includegraphics[width=0.41\textwidth]{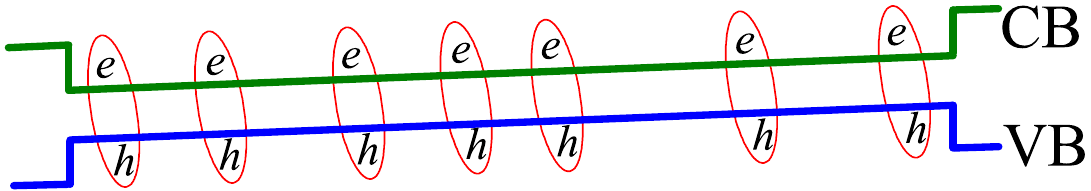}
\caption{(Color online) Spatially separated $e$ and $h$ in declined semiconductor layer (VB is the valence band and CB is the conduction band).}
\vskip -4mm
\end{figure}

The roton-maxon character of the Bogoliubov spectrum \cite{Shlyapnikov, AKSL, pra073031602} is typical for strongly correlated systems \cite{rotonmaxon}, and it was originally observed in liquid helium \cite{He}.
Interesting phenomena and novel many-body phases, {\it e.g.}, crystallization \cite{strcorr} and supersolid \cite{SS}, are realizable in the strong correlation regime for 2D dipolar excitons.

The roton-maxon excitation spectrum and the roton instability effect, as we demonstrate in our work, could be observed as well for a {\it weakly} correlated gas of excitons. 
However, the nature of these phenomena is completely different. 
In the weak correlation regime, the reasons are the anisotropy and the region of attraction of the dipole-dipole interaction potential, which are especially interesting in a semiconductor layer (SL). 
If SL width is greater than several interexcitons distances, exictions with dipole moment, which aligned ``head-to-tail'' along the normal to the plane of SL, start to attract each other. 
As a result of the attraction, initially homogeneous exciton gas becomes unstable under certain conditions, and the system configuration transforms into a spatially inhomogeneous one. 

In the present work, we demonstrate the roton-maxon character of an excitation spectrum for diluted Bose gas of dipolar excitons in SL (see Fig.~1). 
This effect is the result of the region of attraction and the anisotropy of the dipole-dipole interaction in SLs, and it can be illustrated as the residual phenomena of the phonon collapse for dipoles in the extended three-dimensional (3D) system.
We find the threshold for the formation of the roton instability.
We suggest experimental realization for observation of these phenomena. 

The paper is organized as follows.
In Section \ref{vs}, we explain qualitatively the nature of the roton instability for dipolar excitons in SL.
In Section \ref{diag}, we calculate an excitation spectrum and stability diagram. 
In Section \ref{realization}, we find parameters of the system for experimental observation of the roton phenomena in GaAs SL.
Finally, we give our conclusion in Section \ref{Conclusion}.

\section{Excitons in a layer: 2D {\it vs.} 3D}\label{vs}

The stability criterion is the non-negativity of square of the Bogoliubov spectrum \cite{criteria,criteria2},
\begin{eqnarray}\label{ep1}
	\varepsilon_{\vec{\rm p}}^2&=&\frac{{\rm p}^4}{4m^2}+V(\vec{\rm p}\,)\frac{n_0}m{\rm p}^2\geq0, 
\end{eqnarray}
where $\vec{\rm p}$ is the momentum, 
$m$ is the excitons mass, 
$n_0$ is the condensate density,
and $V(\vec{\rm p}\,)$ is the Fourier transform of an interaction potential $V(\vec{\rm r}\,)$.
Here, $\vec{\rm p}=\{{\bf p},p_z\}$ and $\vec{\rm r}=\{{\bf r},z\}$ are 3D vectors, 
${\bf p}$ and ${\bf r}$ are 2D vectors (in the SL plane),
$p=|{\bf p}|$, ${\rm p}=|\vec{\rm p}\,|=\sqrt{p^2+p_z^2}$, and $r=|{\bf r}|$, ${\rm r}=|\vec{\rm r}\,|=\sqrt{{\bf r}^2+z^2}$

Let us consider the Fourier transform $V_d(\vec{\rm p}\,)$ of the dipole-dipole interaction potential $V_d(\vec{\rm r}\,)$ of excitons in the following form:
\begin{equation}\label{Vdp}
	V_d(\vec{\rm p}\,)=\frac{4\pi}3\frac{d^2}{\epsilon}\frac{2p_z^2-p^2}{{\rm p}^2},
	\qquad
	V_d(\vec{\rm r}\,)=\frac{d^2}{\epsilon}\frac{r^2-2z^2}{{\rm r}^5}, 
\end{equation}
where $d=eD$ is the dipole moment of exciton, 
$e>0$ is the hole charge, 
$D$ is the effective $e$-$h$ separation, and 
$\epsilon$ is the dielectric constant.

The key point is the {\it negativity} of the dipole-dipole potential $V_d(\vec{\rm p}\,)$ for the momenta $|p_z|\ll p$ ({\it i.e.}, at $|z|\gg r$), which is a result of the dipole-dipole ``head-to-tail'' attraction.
Thus, for the 3D homogeneous system at $|p_z|\ll p$, the square of the Bogoliubov spectrum (\ref{ep1}) 
\begin{eqnarray}\label{ep21}
	\varepsilon_p^2&=&\frac{p^4}{4m^2}-\frac{4\pi}3\frac{d^2}{\epsilon}\frac{n_0}mp^2 
\end{eqnarray}
is negative for low momenta (see Fig.~2). 
In the phonon region, the spectrum possesses a region of imaginary energies, {\it i.e.}, in the 3D system, phonon modes are unstable.
This regime is known as the phonon instability with respect to the long-wavelength collapse \cite{Shlyapnikov2}. 

\begin{figure}
\includegraphics[width=0.39\textwidth]{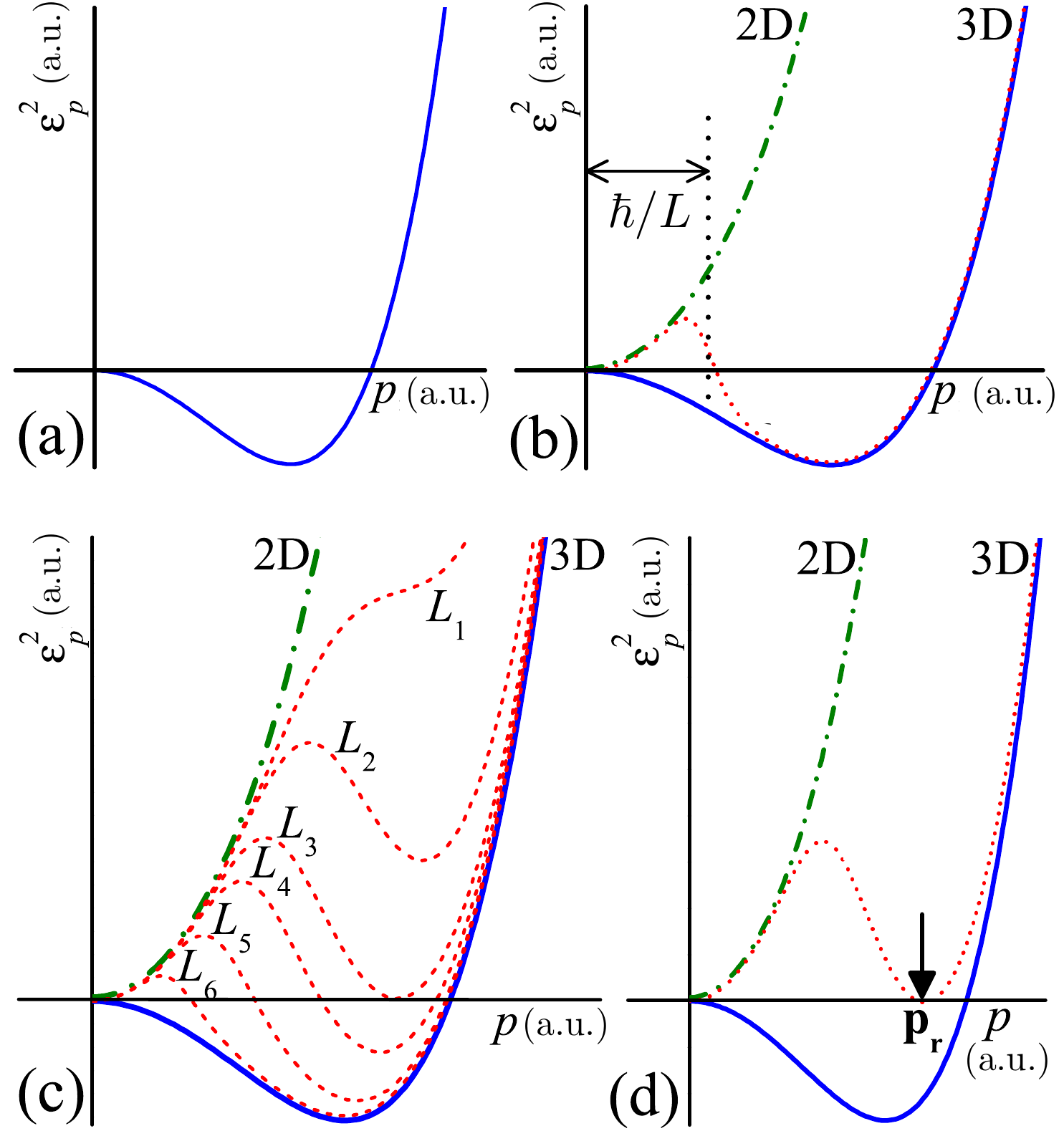}
\caption
{(Color online) Qualitative illustration of appearance of the roton instability for dipolar system in the layer: 
The smooth crossover from 3D (unstable) branch to the 2D (stable) branch
[quantitative examples are presented in Fig. 3a]. 
In (a) and (b) the square of the Bogoliubov spectrum (\ref{ep1}) of dipolar excitons in 3D (solid), 2D (dot-dashed), and the layer geometry (dashed). 
In (c) the width of the SL $L_1{<}L_2{<}\dots{<}L_6$. 
In (d) momentum ${\bf p}_{\rm \,r}$ of the unstable mode.}
\end{figure}

In contrast, let us consider the system of dipolar excitons restricted in $Oz$ direction,
\begin{equation}
	0\le|x|,\!|y|<\infty,
	\qquad 
	0<z<L,
\end{equation}
where $L$ is the width of SL. 
On short-wavelength scales, the motion of excitons corresponds to the 3D regime,
\begin{equation}\label{3D}
	{\rm r}\ll L,
	\qquad 
	{\rm p}\gg\hbar/L. 
\end{equation}
However, on long-wavelength scales ({\it i.e.}, when SL is similar to a thin layer), the motion has 2D behavior:
\begin{equation}\label{2D}
	r\gg L,\quad p\ll\hbar/L,\mbox{ or if }r\gg z,\quad p\ll|p_z|.
\end{equation}

It is clear that 2D regime (\ref{2D}) can be realized for any value of $L$. 
But for 3D regime, the momenta region ${\rm p}\gg\hbar/L$ (or $L\gg\hbar/{\rm p}$) is possible in principle only in SLs, {\it i.e.}, when the following condition holds for the SL width: 
\begin{equation}\label{Q2D}
	L\gg\xi, 
	\qquad 
	\xi=\hbar/\sqrt{2m\mu}
\end{equation}
Here, $\mu$ is the chemical potential of dipolar excitons and $\xi$ is the healing length. 
This is the case of sufficiently wide SL (with both 2D and 3D regimes), which corresponds to the layer system with sufficiently large number $\mathcal{N}\sim{L^2/\xi^2}$ of the energy levels being occupied.

Nevertheless, how does this impact the stability of the system?
If the momenta are sufficiently large $p\gg\hbar/L$, {\it i.e.},  when the 3D regime is realized, the excitation spectrum (\ref{ep1}) is close to its 3D branch. 
This regime is unstable for $|p_z|\ll p$ [see Fig.~2a]. 
{\it Vice versa}, if the momenta are low, {\it i.e.}, when the 2D regime is realized, the spectrum is close to its 2D branch. 
The 2D branch is stable, because the dipoles repel at sufficitenly large distances. 
Therefore, at the momentum interval $p\sim\hbar/L$, there is a smooth crossover from 3D (unstable) branches to 2D (stable) branches [see Fig.~2(b)].

The width of the instability region is determined by the parameter $L$; therefore, for some critical $L$ [$L=L_3$; see Fig.~2(c)], the instability region collapses to the point.
Thus, if the width $L$ of SL is less than the critical one, then the spectrum can have only the roton minimum. 
The critical value of $L$, at which the roton minimum touches the point with zero energy, corresponds to the threshold of the roton instability for dipoles in SL. 
Immediately after the threshold, at which the roton minimum just crosses zero, the square of spectrum (\ref{ep1}) becomes negative for some mode. 

In summary, the formation of the roton minimum and the roton instability is the result of important features of the dipole-dipole interaction 
--- anisotropy and attraction region --- as well as the layer geometry, which passes through unstable 3D and stable 2D regimes.

\section{Stability diagram}\label{diag}

We consider the stability problem for exciton gas in infinite homogeneous SL using the following assumptions. 

First, a distance at which two excitons can approach each other, which is characterized by the total scattering length of excitons, is sufficiently greater that their Bohr radius. 
In this case, an overlap of wave functions of $e$ and $h$ between different excitons is exponentially suppressed. 
Consequently, the fermionic exchange effects \cite{Schindler} and the composite structure of excitons \cite{prl099176403} are negligible. 
Thus, we can consider excitons as bosons.

Second, the Zeeman splitting for excitons in magnetic fields \cite{prb88195309,jetp11900115} is sufficiently greater that other energy scales in the many-body exciton system.
Then the system occupies only the lowest spin branch. 
Thus, excitons have only one spin degree. 

Taking into account these assumptions, we obtain the Hamiltonian of dipolar exciton gas in the following form:
\begin{equation}\label{H}
\begin{split}
	\hat{\mathcal{H}}=\int\hat\Psi^+(\vec{\rm r}\,)\left(-\frac{\hbar^2}{2m}\Delta+V(z)-\mu\right)\hat\Psi(\vec{\rm r}\,)d\vec{\rm r}\\
	+\frac12\int{\hat\Psi^+(\vec{\rm r}\,)\hat\Psi^+(\vec{\rm r}\,')\,\mathcal{U}(\vec{\rm r}-\vec{\rm r}\,')\hat\Psi(\vec{\rm r}\,')\hat\Psi(\vec{\rm r}\,)d\vec{\rm r}\,'d\vec{\rm r}},
\end{split}
\end{equation}
where $\Delta$ is the 3D Laplace operator, 
$\hat\Psi(\vec{\rm r}\,)$ is the excitons field operator satisfying to the standard Bose commutation relations,
$\mathcal{U}(\vec{\rm r}-\vec{\rm r}\,')=V_d(\vec{\rm r}-\vec{\rm r}\,')+{g_s}\delta(\vec{\rm r}-\vec{\rm r}\,')$ is the interaction potential, 
with ${g_s}\delta(\vec{\rm r}\,)$ being the contact van der Waals interaction pseudopotential, 
and ${g_s}$ is the corresponding coupling constant of excitons (${g_s}>0$ if there are no Feshbach resonances), and
\begin{equation}\label{Vz}
	V(z)=\left\{\begin{array}{ll}0,&0<z<L,\\ \infty,&z<0\mbox{ or }z>L\end{array}\right.
\end{equation}
is the confining potential of SL. 

In the weak correlation regime at $T=0$, we can use the Bogoliubov approximation,
\begin{equation}\label{PpP'}
	\hat\Psi(\vec{\rm r},t)=\psi(z)+\hat\Psi'(\vec{\rm r},t),
	\quad
	\psi(z)\equiv\langle\hat\Psi(\vec{\rm r},t)\rangle.
\end{equation} 
Here, $\langle...\rangle$ denotes averaging over the ground state.
The field operator of the noncondensate fraction is sufficiently small (for details, see Appendix IA). 
It has the following representation in the basis $\{\chi_j(z)\}$ in the tight direction
\begin{equation}\label{fraction}
	\hat{\Psi}'(\vec{\rm r}\,)=\sum_{j=0}^{\infty}{\chi_j(z)\hat\Psi'_j({\bf r})},
\end{equation} 
where $\hat\Psi'({\bf r})$ is the corresponding 2D field operator, and the levels with $j>\mathcal{N}$ are approximately nonpopulated.
The order parameter is $\psi(z)=\sqrt{n_0(z)}>0$, 
and $n_0(z)$ does not depend on ${\bf r}$ in the stable phase. 
It is important to note that the Bogoliubov approximation is applicable for systems in the weak correlation regime in SL, whereas it is not in the case of strongly correlated 2D systems \cite{LY2}.

If condition (\ref{Q2D}) takes place, then the chemical potential of excitons $\mu$ is significantly greater than the energy of the lowest level of transversal quantization $\mathcal{E}_0$ in SL of width $L$,
\begin{equation}\label{TF}
	\mu\gg\mathcal{E}_0, \qquad \mathcal{E}_0={\pi^2\hbar^2}/{(2mL^2)}.
\end{equation}
Moreover, if condition (\ref{TF}) holds, then the Thomas-Fermi (TF) regime is realized in the system \cite{prl076000006}.  
In the TF regime, one can neglect the kinetic term in the Gross-Pitaevskii equation (see Appendix IA).
Hence, the order parameter corresponding to (\ref{Vz}) has the form,
\begin{equation}\label{psiz}
	\psi(z)=\sqrt{n_0(z)}=\left\{\begin{array}{ll}\psi\equiv\sqrt{n_0},&0<z<L,\\ 
	0,&\mbox{ otherwise },\end{array}\right.
\end{equation}
where $n_0=\psi^2$ is the 3D BEC density of excitons in the SL \cite{prl092250401}. 
The chemical potential $\mu$ in approximation (\ref{TF}) is equal to $\mu\approx gn_0$, where $g=g_s+2g_d$ with the dipole-dipole coupling constant in the form $g_d=(4\pi/3)d^2/\epsilon$.
It should be noted that due to geometry of the layer [see Eq. (\ref{Umu}], Appendix IIA), multiplication of $g_d$ by the factor of $2$ appears in the total coupling constant \cite{gd}.

In terms of $g$, we can rewrite the dilution conditions as 
\begin{equation}\label{dilute}
\begin{split}
\begin{aligned}
	\beta&\equiv\sqrt{n_0a^3}\ll1, \qquad a=a_s+2a_d, \\
	a_s&=\frac{g_sm}{4\pi\hbar^2}\sim a^*,\qquad a_d=\frac{g_dm}{4\pi\hbar^2}=\frac{md^2}{3\hbar^2\epsilon}, 
\end{aligned}
\end{split}
\end{equation}
where $\beta$ is the gas parameter, 
$a^*$ is the Bohr exciton radius, 
$a_s$ is the van der Waals scattering length, 
$a_d$ is the dipole-dipole scattering length, and $a$ is the total scattering length.

In our consideration, the condensate behavior is described by order parameter (\ref{psiz}).
In turn, the stability of BEC state of the systems is described by fluctuations related to the noncondensate fraction field operator (\ref{fraction}) (see \cite{prb053009341}).

To identify the threshold of instability, we need to find the excitation spectrum. 
For this purpose, we solve the system of the Bogoliubov--de Gennes equations \cite{prb053009341},
\begin{equation}\label{Bu-Bv}
\begin{split}
	\hat Tu_{\vec{\rm p}}(z)+\hat U(u_{\vec{\rm p}}(z)-v_{\vec{\rm p}}(z))&=\varepsilon_{\vec{\rm p}}u_{\vec{\rm p}}(z), \\ 
	\hat Tv_{\vec{\rm p}}(z)+\hat U(v_{\vec{\rm p}}(z)-u_{\vec{\rm p}}(z))&=-\varepsilon_{\vec{\rm p}}v_{\vec{\rm p}}(z),
\end{split}
\end{equation}
where introduced operators $\hat T$ and $\hat U$ are
\begin{equation}\label{TU}
\begin{split}
	\hat T=-\frac{\hbar^2}{2m}\frac{d^2}{dz^2}+\frac{p^2}{2m}+gn_0-\mu, 
\end{split}	
\end{equation}
\begin{equation}\label{TU2}
\begin{split}
	\hat Uf(z)=n_0\times \qquad\qquad \\
	\times{\int_0^L}\left(g\delta(z-z')-\frac{3g_dp}{2\hbar}e^{-p|z-z'|/\hbar}\right)&f(z')dz', 
\end{split}
\end{equation}
with the boundary conditions being 
\begin{equation}\label{boundary}
	u_{\vec{\rm p}}(0)=v_{\vec{\rm p}}(0)=u_{\vec{\rm p}}(L)=v_{\vec{\rm p}}(L)=0.
\end{equation}
Here, we use the notation $p_z=(\pi\hbar/L)l,\; l\in\mathbb{Z_{+}}$.

\begin{figure}
\includegraphics[width=0.5\textwidth]{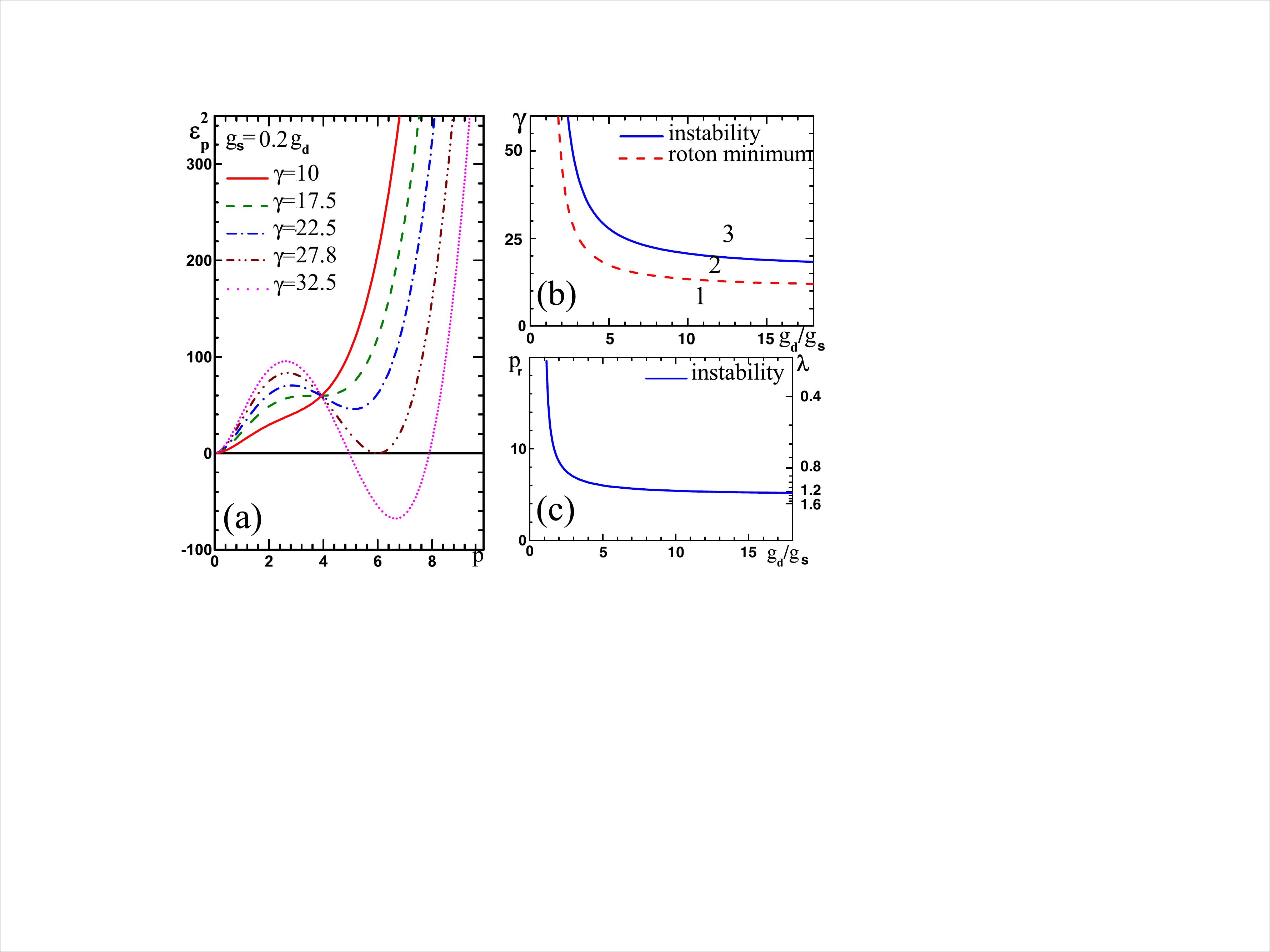}
\vskip -4mm
\caption{
(Color online)
(a) Calculation of the lowest branches of $\varepsilon_p^2$ at 
$g_s=0.2g_d$ and 
$\gamma=10$ (solid), 
$\gamma=17.5$ (dashed), 
$\gamma=22.5$ (dot-dashed), 
$\gamma=27.8$ (dot-dot-dashed) and 
$\gamma=32.5$ (dotted).
(b) Stability diagram in $g_d/g_s$ -- $\gamma$. 
The stable phase without the roton minimum (1), with the roton minimum (2), and the unstable phase (3) are shown.
(c) Critical value of momentum for threshold of instability $p_{\rm \,r}$ and roton wavelength $\lambda=2\pi/p_{\rm \,r}$.
Units $\hbar=m=L=1$ are used in (a)--(c).}
\end{figure}

Bogoliubov-de Gennes equations [Eq. (\ref{Bu-Bv})] are obtained from the Heisenberg equation for the noncondensate fraction field operator (\ref{fraction}) via the Bogoliubov transformation (see Appendix IB).

Solving of Bogoliubov--de Gennes system (\ref{Bu-Bv}) is equivalent to the extremum problem for the following functional:
\begin{equation}\label{Iupvp}
\begin{split}
	I[u_{\vec{\rm p}},v_{\vec{\rm p}}]=\frac12\int_0^L[u_{\vec{\rm p}}\hat Tu_{\vec{\rm p}}+v_{\vec{\rm p}}\hat Tv_{\vec{\rm p}}-\varepsilon_{\vec{\rm p}}(u_{\vec{\rm p}}^2-v_{\vec{\rm p}}^2)+ \\ 
	+(u_{\vec{\rm p}}-v_{\vec{\rm p}})\hat U(u_{\vec{\rm p}}-v_{\vec{\rm p}})]dz+\varepsilon_{\vec{\rm p}}/2
\end{split}
\end{equation}
($z$ dependence of $u_{\vec{\rm p}}$ and $u_{\vec{\rm p}}$ is omitted).

Here, we solve the extremum problem for (\ref{Iupvp}) by the variational method.
In the TF regime (\ref{TF}), we can set $\hat U\approx U=const$ in (\ref{TU2}). 
This dictates the following form for the trial functions (see Appendix IIA):
\begin{equation}\label{upvp}
	u_{\vec{\rm p}}(z)=A\sqrt{\frac2L}\sin\frac{p_zz}{\hbar}, \quad 
	v_{\vec{\rm p}}(z)=B\sqrt{\frac2L}\sin\frac{p_zz}{\hbar},
\end{equation}
with $A,B\ne0$ being the trial parameters. 
After substituting Eq.~(\ref{upvp}) in Eq.~(\ref{Iupvp}), we can find the lowest spectral branch ($p_z=\pi\hbar/L$)
\begin{equation}\label{epb}
	\bar\varepsilon_{\bar p}=\sqrt{\frac{\bar p^4}4+(2+\alpha-A_{\bar p})\gamma\bar p^2},
\end{equation}
whose instability occurs evidently earlier than that for the other branches (see Appendix IIB), with the exciton chemical potential having the explicit form \cite{mu=}
\begin{equation}\label{mu}
	\mu=gn_0+\frac{\pi^2\hbar^2}{2mL^2}.
\end{equation}
In Eq.~(\ref{epb}), we denote $\bar p=pL/\hbar$, $\bar\varepsilon_{\bar p}=mL^2\varepsilon_p/\hbar^2$,
\begin{equation}\label{Apb}
	A_{\bar p}{=}\frac{3\bar p^2}{\bar p^2+\pi^2}{+}\frac{6\pi^2\bar p(1+e^{-\bar p})}{(\bar p^2+\pi^2)^2} \\
	\approx
	\left\{\begin{array}{ll}\mathcal{O}(\bar p),&\bar p\ll\pi,
	\\3&\bar p\gg\pi\end{array}\right.
\end{equation}
\begin{equation}\label{alga}
	\alpha=\frac{g_s}{g_d},
	\qquad
	\gamma=\frac{mg_dn_0L^2}{\hbar^2}=
	\frac{4\pi}{2+\alpha}\left(\frac{\beta L}a\right)^{\!\!2},
\end{equation}
with the case of TF regime (\ref{TF}) corresponding to $(\beta L/a)^2\gg1$ because, typically, $\alpha\sim1$ for excitons.

It should be noted that with an increase of the density $n_0$ and/or the dipole-dipole coupling constant $g_d$ and/or the width $L$ of SL, the spectrum $\varepsilon_p$ bends down [see Fig.~3(a)]. 
As a result of this bending, the roton minimum is formed, which then touches zero, and the homogeneous phase becomes unstable.
The stability diagram is presented in Figs.~3(b) and 3(c).

The phase boundaries for the formation of the roton minimum and the roton instability are determined from the following conditions, respectively:
\begin{equation}\label{rmri}
	\frac{d\varepsilon_p^2}{dp}=\frac{d^2\varepsilon_p^2}{dp^2}=0,\qquad
	\varepsilon_p^2=\frac{d\varepsilon_p^2}{dp}=0.
\end{equation}

From a simple analysis of Eqs. (\ref{epb})--(\ref{alga}) and the stability diagram [Fig.~3(b) and 3(c)], we can conclude that there is no phonon instability in contrast to the 3D case: 
Instability in the system is always the roton instability. 
Subsequently, the unstable phase is possible only if $\alpha<1$, {\it i.e.}, when $g_d>g_s$.
For purely dipolar interaction ($g_s=0$) in sufficiently wide SL (as $\gamma$ is large), the system is always unstable.
The roton minimum and the roton instability are indeed possible only in the TF regime, {\it i.e.}, in the layer geometry rather than in the 2D case \cite{(iv)}.
Furthermore, in the limit of wide SL ($L\to\infty$), for the threshold of the instability we have
\begin{equation}
	g_d-g_s\approx\frac{\pi\hbar}L\sqrt{\frac{3g_d}{mn_0}}=\mathcal{O}\left(\frac1L\right).
\end{equation}

An important question is about the depletion of exciton condensate at $T=0$
\begin{equation}\label{n-n00}
	\frac{n-n_0}{n_0}=\frac1{N_0}\int\langle\hat\Psi'{}^+(\vec{\rm r}\,)
	\hat\Psi'(\vec{\rm r}\,)\rangle d\vec{\rm r}=
	\frac1{N_0}\sum_{\vec{\rm p}}N_{\vec{\rm p}},
\end{equation}
where the summation on ${\vec{\rm p}}$ excludes the term with $l=1$ and ${\bf p}=0$,
$N_{\vec{\rm p}}=\int_0^L|v_{\vec{\rm p}}(z)|^2dz$ is the occupation number of $\vec{\rm p}$ mode,
$N_0=n_0SL$ is the condensate number, and $S$ is the quantization area, which is sufficiently large.
Here, we take into account both in-plane and $Oz$-direction fluctuations.

In our consideration, the lowest spectrum branch corresponds to $l=1$.
Other spectrum branches do not give a divergent contribution (see Appendix IIB).
Close to the threshold of the instability, the lowest spectrum branch $\varepsilon_p$ is close to zero at the roton momentum $p\approx p_{\rm \,r}\ne0$. 
Therefore, the contribution of the lowest branch to the condensate depletion 
\begin{equation}\label{n-n0}
	\frac{n-n_0}{n_0}=\int_0^{\infty}\frac{(\varepsilon_p-p^2/2m)^2}{\varepsilon_pp^2/m}\frac{pdp}{4\pi\hbar^2n_0L}
\end{equation}
diverges logarithmically close to the threshold. 
The divergence of condensate depletion (\ref{n-n0}) indicates that the condensate vanishes close to the threshold of the roton instability, which is in agreement with Ref. \cite{pra073031602}.

However, in sufficiently wide SLs, the following hierarchy between parameters takes place:
\begin{equation}
	a\ll{\xi}\ll{L}\ll\sqrt{S},
\end{equation}
where the first inequality corresponds to dilute regime (\ref{dilute}), the second inequality corresponds to condition of applicability of TF regime (\ref{TF}), and the third one corresponds to the layer geometry.
One can add to this system the following inequality: 
\begin{equation}
	\sqrt{S}\ll{a\exp{\left(\frac{3\beta}{2\pi^{3/2}}\frac{L^2}{\xi^2}\right)}},
\end{equation}
which is satisfied in the deep TF regime [see Eq. (\ref{Q2D})].  
In this case, at $S\to\infty$ the formally divergent term in the noncondensate fraction at sufficiently large $L$,
\begin{equation}\label{divergent}
	\left.\frac{n-n_0}{n_0}\right|^{\rm divergent}_{\rm term}=
	\frac{2\pi^{3/2}}3\frac{\beta}{\gamma}\ln\frac{\beta\sqrt S}{a\gamma}=\mathcal{O}\left(\frac1{L^2}\right)
\end{equation}
can be negligible \cite{diverg}.
Thus, at small $\beta$, in the TF regime (\ref{TF}) the total noncondensate fraction is sufficiently small (see Appendix IIB).

\section{Experimental realization}\label{realization}

We suggest experimental realization of the roton minimum and the roton instability effect for BEC of dipolar excitons in GaAs SL. 
This experimental setup allows one to realize 3D and layer regimes for dipolar excitons with spatially separated $e$ and $h$.

On applying the in-plane magnetic field, the following two advantages could be accomplished.
There is no tunneling dissociation of excitons by the polarizing electric field and the bottom of the TF parabola \cite{prb075035303} shifts from the radiation zones \cite{BH}. 
Thus, for the exciton recombination, a third particle (phonon, impurity) is needed, and as a result, the lifetime becomes sufficient \cite{BH} for cooling down to low temperatures \cite{prl086005608}. 
Besides, in the presence of the in-plane magnetic field, all the above-mentioned calculations are relevant  \cite{prb075035303}.

Details of our estimation: 
the electron mass is $m_e=0.067m_0$ and 
the exciton mass is \cite{prb075035303} $m=0.415m_0$ ($m_0$ is the free electron mass), 
Land\'e $g$ factors \cite{prb048001955} are $g_e,g_h\sim1$, $a^*=11.8$ nm \cite{prb075035303}, 
$\epsilon=12.5$, $a_s=1.7a^*=20$ nm, $a_d=22$ nm, 
the gas parameter $n_0a^3=0.19\ll1$ corresponds to the value $\beta=(3\sqrt{\pi}/2)(n-n_0)/n_0=0.44$ on the threshold of the instability in the limit $L\to\infty$, 
the value $D=\sqrt{3a'a_d/2}=10$ nm (where $a'=\hbar^2\epsilon/me^2=1.59$ nm) corresponds to the fields \cite{prb075035303} $B_{\parallel}=4$ T and $E_{\perp}=1.5$ kV$/$cm, 
the values of parameters $\gamma=(4\pi/3)(\beta L/a)^2=3.2\cdot10^3$ and $\alpha=1-\pi\sqrt{3/\gamma}=0.9$ correspond to the deep TF regime, 
values \cite{BH} $p_{\parallel}=eDB_{\parallel}/c_0$ and $q_r=E_g/c$ correspond to the exciton gap in GaAs is $E_g=1.51$ eV, 
light velocity in GaAs $c=c_0/\sqrt{\epsilon}$ and $c_0$ in vacuum, $c_s=2\sqrt{\pi}\hbar\beta/ma$,
$\lambda=\sqrt[\mbox4]{3\gamma/4}a/\beta$, and 
we can neglect the anisotropy of $m$ and its dependence from $B_{\parallel}$ and $E_{\perp}$.

In our estimations, we suggest that $L=4$ $\mu$m, $\lambda=1$ $\mu$m, $n_0/n=6/7$, normal electric field is $E_{\perp}=1.5$ kV/cm and in-plane magnetic field is $B_{\parallel}=4$ T. 
Under these conditions, at the threshold of the roton instability, exciton BEC in GaAs SL's is realizable. 

This is justified by the following arguments:

(i) The momentum displacement of the exciton parabola bottom $p_{\parallel}=6.1\cdot10^5$ cm$^{-1}\hbar$ is sufficiently greater than the radiation zone width $q_r=2.7\cdot10^5$ cm$^{-1}\hbar$ in GaAs. 
Thus, excitons are actually ``dark", {\it i.e.}, they are long-lived.

(ii) Zero-sound velocity in the exciton system $c_s=6.8\cdot10^5$ cm/s is greater than the velocity of longitudinal sound in GaAs $c_{\rm phon}=5.36\cdot10^5$ cm$/$s. 
This provides efficient cooling of excitons by the GaAs lattice.

(iii) The chemical potential of excitons $\mu=1.27$ K is sufficiently lower than the Zeeman splitting $\Delta E\sim5$ K \cite{prb88195309,jetp11900115}. 
Therefore, at low temperatures, gas of excitons has only one spin degree.

(iv) Under these conditions, in the regime of spatially separated cw pump \cite{njp014105007, prb083165308, prl096227402} and the evaporative cooling \cite{prl103087403}, 
the artificially trapped \cite{B,prb083165308,prl103087403,prl096227402,S,nl0012000326} excitons are readily cooled to very low temperatures \cite{prb088041201}, 
which are still lower than the temperature for BEC in 3D ideal gas $T_{\rm BEC}^{\rm IG}=650$ mK.

(v) The scattering length of excitons $a=64$ nm is sufficiently greater than their Bohr radius $a^*=11.8$ nm. 
Therefore, the tunnel transformation of excitons into biexcitons \cite{prb087035302} is suppressed. 
Besides, destruction of BEC \cite{Maezono} and superfluidity \cite{pssc00302457} by the Fermi exchange effects is exponentially small \cite{jetpl1997}.

(vi) The exciton density $n=0.86\cdot10^{15}$ cm$^{-3}$ is higher by two orders than the concentration of impurities in pure GaAs samples.
Hence, the influence of the latters is negligible. 
Moreover, free carriers \cite{trions} can be compensated \cite{K} by the spatially indirect injection \cite{jetpl0940800}. 
Lastly, two-exciton recombination processes were not observed in GaAs heterostructures \cite{prb073035205}.

We note that condition (iii) corresponds to the system of single spin component dipolar excitons.
In other words, predicted effects are valid for the lowest spin branch. 
However, presence of the exchange interaction between excitons could be a reason for a set of interesting collective phenomena, 
including phase transitions between instabilities in different spin components controlled by the detuning of the magnetic field \cite{prl109026401}. 
The roton phenomena for multispin component exciton system will be considered in detail in another place.

\section{Conclusions}\label{Conclusion}

In the present work, we have considered the system of dipolar excitons with one spin degree in the weak correlation regime in the layer geometry. 
For the system in SL, we have predicted the roton-maxon character of the excitation spectrum and the roton instability effect. 
For the experimental verification of these effects, we have suggested realistic experimental realization for SL in GaAs heterostructures. 

However, it should be noted that in this work we have focused on the case of isotropic $h$ mass. 
The anisotropy of $h$ mass provides additional symmetry breaking in the system, which can be a reason for interesting structural properties, {\it e.g.}, density waves phases at the threshold of the roton instability. 
To precisely determine the ground state of the system with respect to this effect, additional calculations are needed. 

\section*{Acknowledgements}

We thank E.A. Demler, Y.E. Shchadilova, and V.I. Yudson for fruitful discussions
as well as O.V. Kotov, D.V. Kuznetsov, and O.V. Lychkovskiy for clarifying discussions.
The work is supported by the RFBR (14-02-00937 \& 14-08-00606) and the RAS programs. 
A.K.F. is supported by the Dynasty Foundation.
Yu.E.L. is supported by the HSE Program of Basic Research. 

\setcounter{equation}{0}
\renewcommand{\theequation}{A\arabic{equation}}

\section*{Appendix I. Bogoliubov theory for dipolar excitons in a layer}
\subsection{Order parameter in the TF approximation}
Consider the Heisenberg equation for the exciton field with Hamiltonian (\ref{H}) at $0<z<L$
\begin{equation}\label{UG}
\begin{split}
	i\hbar\frac{\partial\hat\Psi(\vec{\rm r},t)}{\partial t}=\left(-\frac{\hbar^2}{2m}\Delta-\mu\right)\hat\Psi(\vec{\rm r},t) \\
	+\int\mathcal{U}(\vec{\rm r}-\vec{\rm r}\,') 
	\hat\Psi^+(\vec{\rm r}\,',t)\hat\Psi(\vec{\rm r}\,',t)
	d\vec{\rm r}\,'\hat\Psi(\vec{\rm r},t).
\end{split}
\end{equation}
In the weak correlation regime, we can use the Bogoliubov approximation $\psi(z)\equiv\langle\hat\Psi(\vec{\rm r},t)\rangle$ (\ref{PpP'}).

We stress the conceptually important case of zero temperature $T=0$. 
In this case, in the weak interaction regime, the total density of excitons $n$ is close to the density of the exciton condensate $n\approx n_0$, and the noncondensate fraction is small $n-n_0\ll n_0$.
Therefore, the corresponding field operator $\hat\Psi'(\vec{\rm r},t)$ is small:
\begin{equation}\label{P'p}
	||\hat\Psi'(\vec{\rm r},t)||\ll||\psi(z)||,
\end{equation}
where we use the notations
\begin{eqnarray}
	||\hat\Psi'(\vec{\rm r},t)||^2&=&{\int}{d\vec{\rm r}\,\langle{\hat\Psi'^{+}(\vec{\rm r},t)\hat\Psi'(\vec{\rm r},t)}\rangle}, \\
	||\psi(z)||^2&=&{\int}{d\vec{\rm r}\,\psi^2(z)}  
\end{eqnarray}

Thus, we can linearize the exciton field product in (\ref{UG}),
\begin{equation}\label{PPP}
\begin{split}
	&\hat\Psi^+(\vec{\rm r}\,',t)\hat\Psi(\vec{\rm r}\,',t)\hat\Psi(\vec{\rm r},t)=\psi(z)\psi^2(z')+ \\ 
	&\psi^2(z')\hat\Psi'(\vec{\rm r},t)+\psi(z)\psi(z')\{\hat\Psi'(\vec{\rm r}\,',t)+\hat\Psi'{}^+(\vec{\rm r}\,',t)\}. 
\end{split}
\end{equation}

By substituting (\ref{psiz}) in (\ref{PPP}) and (\ref{PPP}) in (\ref{UG}), with averaging on the ground state, we obtain
\begin{equation}\label{Umu}
\begin{split}
	\psi^3{\int}{d{\bf r}}'{\int_0^L dz'\mathcal{U}(\vec{\rm r}-\vec{\rm r}\,')}=g\psi^3=\mu\psi,
\end{split}
\end{equation}
It should be noted from Eq.~(\ref{Umu}) that the value $g$ has a sense of the coupling constant for excitons in the case of both dipole-dipole and van der Waals interactions. 

From Eq. (\ref{Umu}), we can find the chemical potential of excitons in the following form:
\begin{equation}\label{muO}
	\mu=gn_0+\mathcal{O}\left(\mathcal{E}_0\right).
\end{equation}
One can see, that the TF approximation has error $\mathcal{O}\left(\mathcal{E}_0\right)$.
This fact is in agreement with relation (\ref{TF}).  
This error appears as a result of the bending of the order parameter $\psi(z)$ to zero near the boundary of SL $\{z=0,L\}$, which is ignored in the TF approximation \cite{prl076000006}.

\subsection{Bogoliubov--de Gennes equations}
By substituting (\ref{psiz}) in (\ref{PPP}), (\ref{PPP}) in (\ref{UG}) and by taking (\ref{Umu}) into account, after transformations, we get the following equation for the noncondensate field operator ($0<z<L$):
\begin{equation}\label{UP'}
\begin{split}
	i\hbar\frac{\partial\hat\Psi'(\vec{\rm r},t)}{\partial t}&=\left(-\frac{\hbar^2}{2m}\Delta+gn_0-\mu\right)\hat\Psi'(\vec{\rm r},t) \\
	+n_0{\int}d{\bf r}'{\int_0^L}&dz'\mathcal{U}(\vec{\rm r}-\vec{\rm r}\,')\left(\hat\Psi'(\vec{\rm r}\,',t){+}\hat\Psi'{}^+(\vec{\rm r}\,',t)\right).
\end{split}
\end{equation}

To solve Eq.~(\ref{UP'}), we use the Bogoliubov transformation,
\begin{equation}\label{PsiB}
\begin{split}
	&\hat\Psi'(\vec{\rm r}\,)=\frac1{\sqrt S}\sum\nolimits_{\vec{\rm p}}e^{\frac i{\hbar}{\bf pr}}\left(u_{\vec{\rm p}}(z)\hat a_{\vec{\rm p}}-v_{\vec{\rm p}}(z)\hat a_{-\vec{\rm p}}^+\right),\\
	&\hat a_{\vec{\rm p}}=\int\frac{dzd{\bf r}}{\sqrt S}e^{-\frac i{\hbar}{\bf pr}}\left(u_{\vec{\rm p}}(z)\hat\Psi'(\vec{\rm r}\,)+v_{\vec{\rm p}}(z)\hat\Psi'{}^+(\vec{\rm r}\,)\right).
\end{split}
\end{equation}
Here, $\hat a_{\vec{\rm p}}$ and $\hat a^+_{\vec{\rm p}}$ are the operators of Bogoliubov excitations satisfying the standard Bose commutation relations,
\begin{equation}
	[\hat a_{\vec{\rm p}},\hat a_{\vec{\rm p}'}]=0, \qquad [\hat a_{\vec{\rm p}},\hat a_{\vec{\rm p}'}^+]=\delta_{\vec{\rm p}\vec{\rm p}'},
\end{equation}
$u_{\vec{\rm p}}(z)$ and $v_{\vec{\rm p}}(z)$ are the Bogoliubov $uv$-functions, which satisfy (\ref{Bu-Bv})
with the boundary conditions (\ref{boundary}), the conditions of the normalization,
\begin{equation}\label{PsiB}
\begin{split}
\begin{aligned}
	&{\int}\frac{d{\bf r}}{S}{\int_0^L}\!\!dze^{-\frac i{\hbar}({\bf p}{-}{\bf p'}){\bf r}}(u_{\vec{\rm p}}(z)u_{\vec{\rm p}'}(z){-}v_{\vec{\rm p}}(z)v_{\vec{\rm p}'}(z)){=}\delta_{\vec{\rm p}\vec{\rm p}'}, \\ 
	&{\int}\frac{d{\bf r}}{S}\!\!\int_0^L\!\!dze^{-\frac i{\hbar}({\bf p}{-}{\bf p'}){\bf r}}(u_{\vec{\rm p}}(z)v_{\vec{\rm p}'}(z){-}v_{\vec{\rm p}}(z)u_{\vec{\rm p}'}(z)){=}0, 
\end{aligned}
\end{split}
\end{equation}
and the completeness,
\begin{equation}
\begin{split}
\begin{aligned}
	&\frac1S\sum\nolimits_{\vec{\rm p}}e^{\frac i{\hbar}{\bf p}({\bf r}-{\bf r}')}(u_{\vec{\rm p}}(z)u_{\vec{\rm p}}(z')-v_{\vec{\rm p}}(z)v_{\vec{\rm p}}(z'))=\delta(\vec{\rm r}-\vec{\rm r}\,'), \\ 
	&\frac1S\sum\nolimits_{\vec{\rm p}}e^{\frac i{\hbar}{\bf p}({\bf r}-{\bf r}')}(u_{\vec{\rm p}}(z)v_{\vec{\rm p}}(z')-v_{\vec{\rm p}}(z)u_{\vec{\rm p}}(z'))=0. 
\end{aligned}
\end{split}
\end{equation}
After Bogoliubov $uv$ transformation (\ref{PsiB}), Eq. (\ref{UP'}) reduces to the simple equation for the Heisenberg operator $\hat a_{\vec{\rm p}}(t)$
\begin{equation}
	i\hbar\frac d{dt}\hat a_{\vec{\rm p}}(t)=\varepsilon_{\vec{\rm p}}\hat a_{\vec{\rm p}}(t), \quad \hat a_{\vec{\rm p}}(t)=e^{-\frac i{\hbar}\varepsilon_{\vec{\rm p}}t}\hat a_{\vec{\rm p}},
\end{equation}
and system of Bogoliubov-de Gennes equations (\ref{Bu-Bv}).

For the calculation of the Bogoliubov spectrum $\varepsilon_{\vec{\rm p}}$, we need to solve system (\ref{Bu-Bv}).
It is clear that system (\ref{Bu-Bv}) is obtained from the extremum problem for functional (\ref{Iupvp}).
From the conditions ${\delta I}/{\delta u_{\vec{\rm p}}}=0$ and ${\delta I}/{\delta v_{\vec{\rm p}}}=0$,
we can obtain equations (\ref{Bu-Bv}), and by minimization of $I$ on $\varepsilon_p$, we can get the normalization condition at $\vec{\rm p}=\vec{\rm p}'$.

\section*{Appendix II. Variational approach}
\subsection*{A. Trial functions}
We need to choose the form of the trial functions $u_{\vec{\rm p}}(z)$ and $v_{\vec{\rm p}}(z)$ for functional (\ref{Iupvp}). 
We note that in the TF regime, the typical momentum of the system ${\rm p}\sim\hbar/\xi$ satisfies the inequality ${\rm p}\gg\pi\hbar/L$. 
However, the roton instability first appears for the lowest spectral branch $l=1$ with $p_z=\pi\hbar/L$.
Thus, we have  
\begin{equation}
	p=\sqrt{{\rm p}^2-p_z^2}\approx{\rm p}\gg\pi\hbar/L. 
\end{equation}

Therefore, the typical scale of change in (\ref{TU}) is estimated as $|z-z'|\ll L/\pi$. 
At the same time, the Bogoliubov modes $u_{\vec{\rm p}}(z)$ and $v_{\vec{\rm p}}(z)$ for the spectral branch, obviously, changes on the scale $\approx{L}$, which is greater. 
As a result, we can substitute the exponential part $e^{-p|z-z'|/\hbar}$ in (\ref{TU}) to obtain 
\begin{equation}
	(p/\hbar)e^{-p|z-z'|/\hbar}\approx2\delta(z-z').
\end{equation}
The operator $\hat U\approx U=const$ is local.

Therefore, the trial functions $u_{\vec{\rm p}}(z)$ and $v_{\vec{\rm p}}(z)$ are useful in form (\ref{upvp}).

\subsection*{B. Variational method and condensate depletion}
The extremum for the functional $I(A,B)$ [see Eq. (\ref{Iupvp})] for nontrivial values of $A$ and $B$ ($A,B\ne0$) is achieved if the condition,
\begin{equation}\label{ep0}
	\varepsilon_{\vec{\rm p}}=\sqrt{C_1(\vec{\rm p})(C_1(\vec{\rm p})+2C_2(\vec{\rm p}))}, 
\end{equation}
holds, and the values of $A$ and $B$ in the extremum point have the form,
\begin{equation}\label{AB}
	A^2=\frac{(\varepsilon_{\vec{\rm p}}+C_1(\vec{\rm p}))^2}{4\varepsilon_{\vec{\rm p}}C_1(\vec{\rm p})}, \quad
	B^2=\frac{(\varepsilon_{\vec{\rm p}}-C_1(\vec{\rm p}))^2}{4\varepsilon_{\vec{\rm p}}C_1(\vec{\rm p})},
\end{equation}
where $A$ and $B$ are normalized in view of the normalization conditions.
Here,
\begin{equation}\label{c1c2}
\begin{split}
\begin{aligned}
	&C_1(\vec{\rm p})={\rm p}^2/2m+gn_0-\mu, \\
	&C_2(\vec{\rm p})=gn_0-\frac{3g_dn_0p}{\hbar L}C(\vec{\rm p}),
\end{aligned}
\end{split}
\end{equation}
\begin{equation}\label{cl}
\begin{split}
\begin{aligned}
	C(\vec{\rm p})&={\int}{\int}_0^L{\left(\sin\frac{p_zz}{\hbar}\sin\frac{p_zz'}{\hbar}e^{-p|z-z'|/\hbar}\right)}dzdz' = \\
	&=\frac{\bar{p}L^2}{\pi^2l^2+\bar{p}\,^2}+\frac{2\pi^2l^2L^2}{\left(\pi^2l^2+\bar{p}\,^2\right)^2}\left(1-(-1)^{l}e^{-\bar{p}}\right)
\end{aligned}
\end{split}
\end{equation}

After substituting (\ref{c1c2}) and (\ref{cl}) at $p_z=\pi\hbar/L$ in (\ref{ep0}) and transformations, we get Eq. (\ref{epb}).
Furthermore, by substituting (\ref{c1c2}) and (\ref{cl}) in (\ref{ep0}), one can prove that (i) the lowest spectrum branch corresponds to $l{=}1$;
(ii) instability for the branch with $l{=}1$ occurs earlier than that for other branches;
(iii) the divergence in the condensate depletion is possible for $l=1$ spectrum branch only.

By substituting (\ref{c1c2})-(\ref{cl}) in (\ref{ep0}) as well as (\ref{ep0}) in (\ref{AB}) and integrating (\ref{AB}) over the momentum space, we obtain Eq. (\ref{n-n00}) for the condensate depletion [see Eq. (\ref{upvp})]. 

We note that in the TF regime (\ref{TF}) in SLs, we can neglect the latter term in (\ref{cl}), because it is small as $\mathcal{O}(1/L)$ at $p,p_z\gg{\pi\hbar/L}$.
In this case at sufficiently large $L$, for Eq. (\ref{n-n0}) we obtain the following expression at the threshold of the roton instability:
\begin{equation}\label{n-n000}
	\frac{n-n_0}{n_0}=\left.\frac{n-n_0}{n_0}\right|_{p,p_z\gg{\pi\hbar/L}}+\frac{2\pi^{3/2}}3\frac{\beta}{\gamma}\ln\frac{\beta\sqrt S}{a\gamma}.
\end{equation}
Here, the first term corresponds to an impact of all spectrum branches, including both in-plane and $Oz$-directions fluctuations, as well as the convergent part of the lowest spectrum branch. 
Due to dilute regime (\ref{dilute}), it is small as ${2\beta}/{3\sqrt{\pi}}$.
The second term, which comes from the formal divergence of the lowest spectrum branch, is small as $\mathcal{O}(1/L^2)$ [see Eq. (\ref{divergent})].


\begin{thebibliography}{}
\bibitem{gd}
Conventionally, we introduce the dipole-dipole coupling constant as $g_d=(4\pi/3)d^2/\epsilon$. 
Due to geometry of the layer [see Eq. (\ref{Umu}) and Appendix IIA], in the total coupling constant $g=g_s+2g_d$, multiplication of $g_d$ by the factor of $2$ appears.
Another way \cite{Shlyapnikov} is to introduce the dipole-dipole coupling constant  as $\tilde{g}_d=(8\pi/3)d^2/\epsilon$, then the total coupling constant becomes $g=g_s+\tilde{g}_d$.
\bibitem{mu=}
This is due to the fact that the lowest branch must vanish at $p\to0$, since stable 2D gas is superfluid at $T=0$.
Thus, using (\ref{muO}), we find explicit relation (\ref{mu}) for the chemical potential of excitons.
\bibitem{(iv)}
In the absence of the Feshbach resonance, the van der Waals interaction of excitons is repulsive: $g_s>0$. 
For this case, our numerical solution of Eq.~(\ref{rmri}) with the lowest branch (\ref{epb})--(\ref{Apb}) shows that the roton minimum exists only if $\mu>5.26\mathcal{E}_0$ and the roton instability for $\mu>7.39\mathcal{E}_0$. 
In both cases $\mu\gg\mathcal{E}_0$, thus, the TF regime (\ref{TF}) is realized indeed.
\bibitem{diverg}
Taking into account the hole-mass anisotropy in GaAs, which results in the anisotropy of exciton mass $m$ and the lowest branch $\varepsilon_p$, 
we can show that the condensate depletion converges.
A similar situation takes place in systems of tilted dipolar atoms \cite{AKSL}.
\end{thebibliography}
\end{document}